\documentclass[a4paper]{jpconf}
\usepackage{graphicx}
\usepackage{amsmath}
\begin{document}
\title{Cold collisions of C$_{2}^{-}$ anions with Li and Rb atoms in hybrid traps}

\author{Milaim Kas$^1$, Jacques Li\'evin$^1$, Nathalie Vaeck$^1$ and J\'er\^ome Loreau$^{1,2}$}
\address{$^1$ Service de Chimie Quantique et Photophysique, Universit\'e libre de Bruxelles CP 160/09, 1050 Brussels, Belgium}
\address{$^2$ KU Leuven, Department of Chemistry, B-3001 Leuven, Belgium}

\begin{abstract}
We present a theoretical investigation of reactive and non-reactive collisions of Li and Rb atoms with C$_{2}^{-}$ molecular anions at low temperatures in the context of sympathetic cooling in hybrid trap experiments. 
Based on recently reported accurate potential energy surfaces for the singlet and triplet states of the Li--C$_{2}^{-}$ and Rb--C$_{2}^{-}$ systems, we show that the associative electronic detachment reaction is slow if the colliding partners are in their ground state, but fast if they are excited. The results are expected to be representative of the alkali-metal series. We also investigate rotationally inelastic collisions in order to explore the cooling of the translational and rotational degrees of freedom of C$_2^-$ in hybrid ion-atom traps. The effect of micromotion is taken into account by considering Tsallis distributions of collision energies. We show that the translational cooling occurs much more rapidly than rotational cooling and that the presence of excited atoms leads to losses of anions on a timescale comparable to that of rotational cooling.

\end{abstract}

\section{Introduction}
Theoretical and experimental studies of low temperature reactive and non-reactive collisions involving molecular anions have drawn interest recently, both due to their importance in astrochemistry \cite{Millar2017a} as well as in the field of cold physics and chemistry \cite{Tomza2019_review}. In this context, one of the main experimental objectives is to produce samples of (ultra)cold molecular anions. 
While direct laser cooling anions is challenging \cite{Cerchiari2019}, these species can be trapped efficiently using either radio-frequency (rf) or Penning traps. By co-trapping anions with ultracold atoms, it should therefore be feasible to sympathetically cool down anions by collisions.
In order to investigate this possibility, a knowledge of the collisional properties of anions at low temperature is particularly important. In the case of anions the main reactive channel leading to losses from the trap is usually the associative electronic detachment reaction, A~+~BC$^{-}$~$\rightarrow$~ABC~+~$e^{-}$.
This has been observed experimentally for collisions of ultracold Rb atoms with OH$^-$ in a hybrid trap composed of a magneto-optical trap (MOT) combined with an octupole rf trap \cite{Deiglmayr2012} and explained on the basis of accurate \textit{ab initio} calculations \cite{Kas2016}. In particular, we have recently shown that among the alkali and alkali earth series, the associative electronic detachment of OH$^-$ at low temperature is expected to proceed efficiently only for collisions involving Rb and Cs atoms \cite{Kas2017}. For the anions that are stable against collisional detachment, sympathetic cooling could be feasible.

Besides OH$^-$, other candidates for producing cold molecular anions have been suggested \cite{Tomza2017}. An interesting species that has been widely studied both theoretically and experimentally is C$_2^-$, an open-shell species with $^2\Sigma_g^+$ symmetry and an electron affinity of 3.2 eV. Furthermore, it possesses several electronic excited states that could be used in laser-cooling schemes \cite{Yzombard2015} as well as to explore excited-state dynamics. In a recent work, we have studied the interaction of C$_2^-$ with Li and Rb atoms and concluded that the associative detachment reaction should proceed with a rate much smaller than the Langevin rate provided that the ultracold atoms are in their ground state \cite{Kas2019a}. In section \ref{section_ab_initio} we discuss the potential energy surfaces (PESs) that were constructed for these systems. We then investigate reactive and non-reactive collisions of C$_2^-$ with Li and Rb atoms in Section \ref{section_collisions} and present a model for the translational and rotational cooling of C$_2^-$ in hybrid traps in Section \ref{section_cooling}.

\section{\textit{Ab initio} calculations}\label{section_ab_initio}

\begin{figure}[h]
\centering
\begin{minipage}{20pc}
\hspace{-2pc}%
\includegraphics[width=1\textwidth]{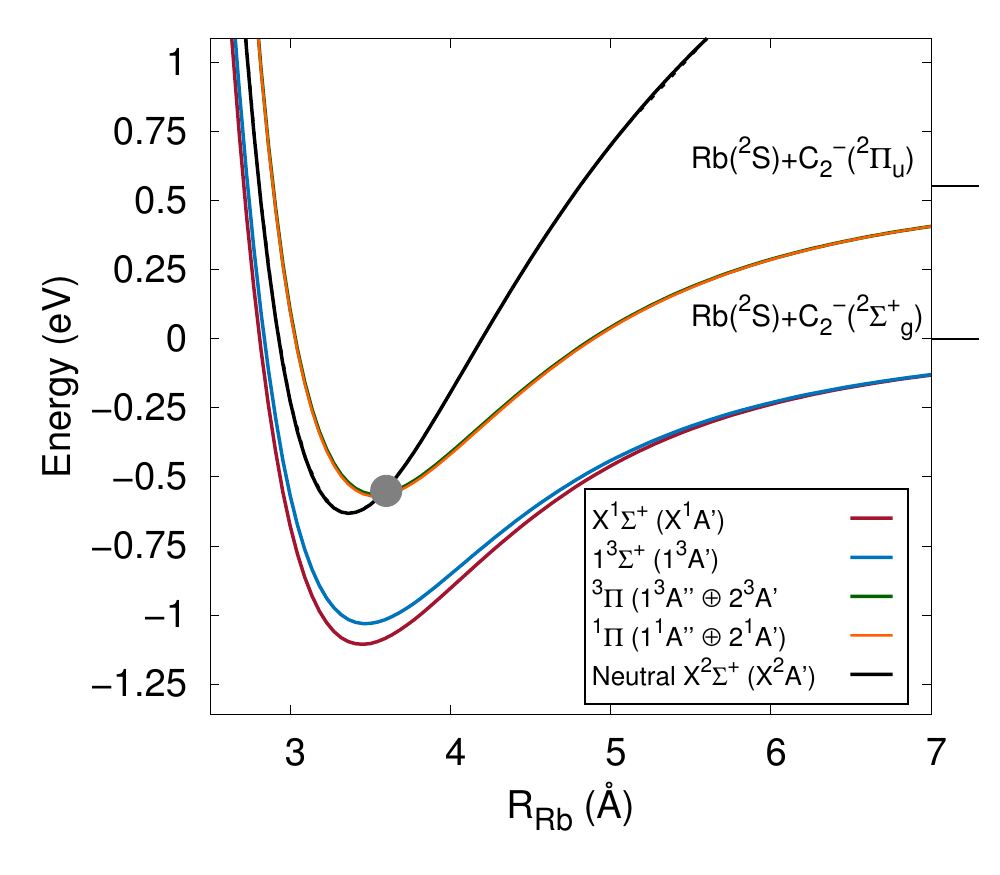}
\end{minipage}%
\begin{minipage}{20pc}
\hspace{-2pc}%
\includegraphics[width=1\textwidth]{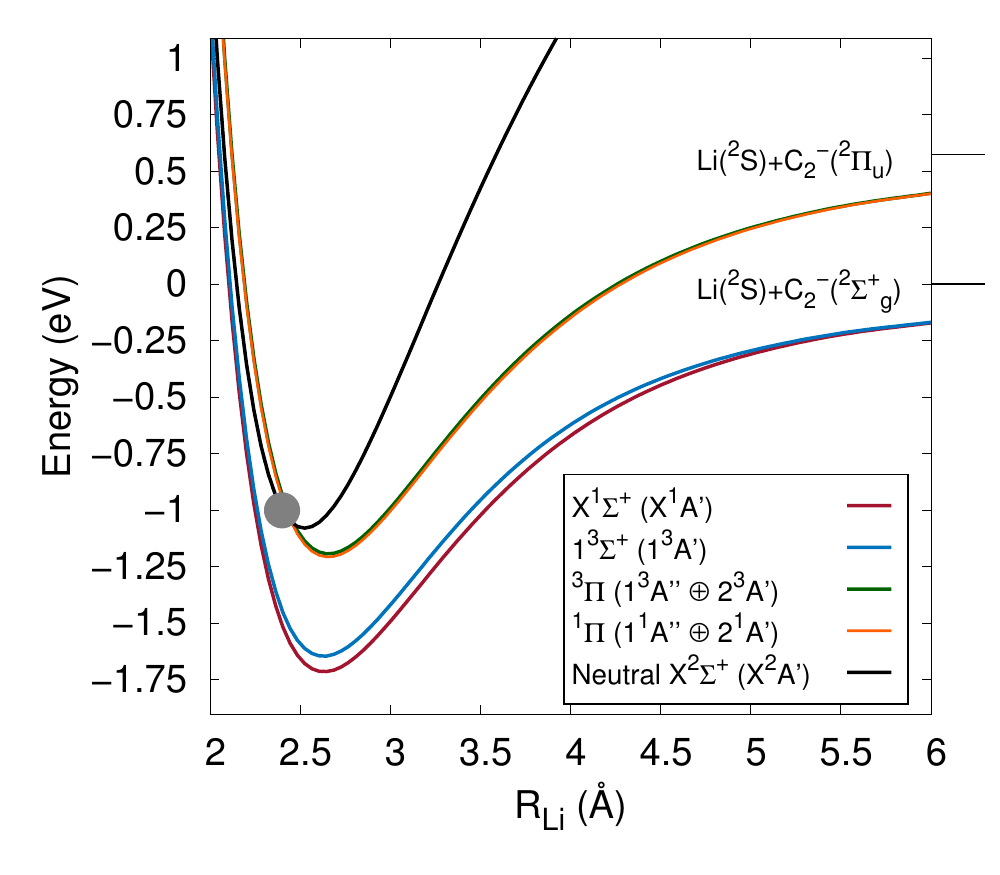}
\end{minipage} 
\vspace{-.5cm}
\caption{\label{fig_PES}
Potential energy curves for the lowest states of Rb--C$_{2}^{-}$ (left) and Li--C$_{2}^{-}$ (right) molecular species at $\theta=0^\circ$. The energy is given relative to the ground state dissociation energy. The two dissociation limits at which the molecular states correlate are shown. The full grey circle shows the crossing geometry between the anion and neutral potential energy curves.}
\end{figure}

Accurate \textit{ab initio} calculations on the Li--C$_2^-$ and Rb--C$_2^-$ systems have been performed with the CASSCF/ic-MRCI+Q method and large basis sets \cite{Kas2019a}. The system is described using the Jacobi coordinates $R$ (the distance of the atom to the center of mass of C$_2^-$) and $\theta$ (the relative angle), while the internuclear distance of C$_2^-$ $r$ is fixed to its equilibrium value, 1.268~\AA. 
The interaction between an alkali atom and C$_{2}^{-}(X\, ^{2}\Sigma^{+}_{g})$ leads to $^{1,3}\Sigma^{+}$ states for linear geometries (C$_{\infty v}$ point group) or $A'$ states at bent geometries (C$_{s}$ point group).  
Two-dimensional PESs were obtained for both systems in singlet and triplet states, the former being the ground state of the collisional complex. The equilibrium geometry corresponds to the T-shaped configuration with a well depth of 2.30 eV for Li-C$_2^-$ and 1.33 eV for Rb-C$_2^-$.
An interesting feature of C$_{2}^{-}$ is that it presents several bound excited electronic states, which opens the possibility of studying excited-state reactivity in anionic systems as well as the use of Doppler thermometry \cite{Gianfrani2016} to measure the translational temperature of C$_2^-$. We obtained PESs for the lowest excited molecular states of the complex, which correlate to the Li/Rb($^2S$)+C$_{2}^{-}$($^{2}\Pi_{u}$) dissociation channel, located 0.50 eV above the ground state.
Finally, we have also calculated the PESs for the corresponding neutral species (Li--C$_2$ and Rb--C$_2$) in order to investigate the associative detachment reaction. The neutral ground state ($X\ ^2\Sigma^+_g$ or $X\ ^2A'$) was incorporated along with the anionic states into the state-averaged procedure in order to obtain a common CASSCF wave function. 
The results of these \textit{ab initio} calculations are shown in Figure \ref{fig_PES} for Li--C$_2^-$ and Rb--C$_2^-$ for the linear geometry ($\theta=0^\circ$).

\section{Collisional dynamics}\label{section_collisions}
\subsection{Reactive Collisions: associative electron detachment}\label{section_reactive}

The associative electron detachment (AED) process A + BC$^{-} \rightarrow$ ABC + $e^{-}$, in which a bond is formed while an electron is ejected, can proceed through two different mechanisms. 
The fastest one occurs when the system reaches a geometry at which the neutral and anionic PESs of the complex cross. The range of nuclear coordinates for which the anion PES lies higher in energy than the neutral PES defines the autodetachment region where the wave function associated to the entrance channel is embedded into the electronic continuum. At such geometries, the anionic state is unstable and the excess electron is spontaneously ejected. The rate at which electron detachment occurs depends on the probability that the electron tunnels through a centrifugal barrier and will therefore be larger when the excess electron occupies an orbital with a small orbital quantum number. This can be calculated using model potentials or more elaborated methods, such as R-matrix approaches \cite{Tennyson2010}, that allow to directly obtain the width, hence the lifetime, of these autodetaching states. Since it involves electronic motion, the detachment rate is usually very high ($10^{14}-10^{15}$ s$^{-1}$ \cite{Simons1998}). The second mechanism involves an energy transfer between nuclear and electronic motions and occurs through non-adiabatic couplings. 
It was found \cite{Simons1998} that the following characteristics lead to large detachment rates:
\textit{(i)} a very diffuse highest occupied molecular orbital (HOMO) that varies quickly with $R$; \textit{(ii)} a small energy difference between anion and neutral PES; \textit{(iii)} a small reduced mass.
Although the calculation of the detachment rate and is challenging, general trends and approximate rates can be obtained on the basis of the PESs as well as energetic considerations. Since the first mechanism (``curve jumping process'' into the autodetachment region) leads to detachment rates that are usually several orders of magnitude larger than for the second mechanism (non-adiabatic driven electron detachment), one can neglect the latter if the neutral and anion PESs cross in an energetically accessible region of the nuclear coordinates.

\begin{figure}[h!]
\begin{minipage}{27pc}
\includegraphics[width=27pc]{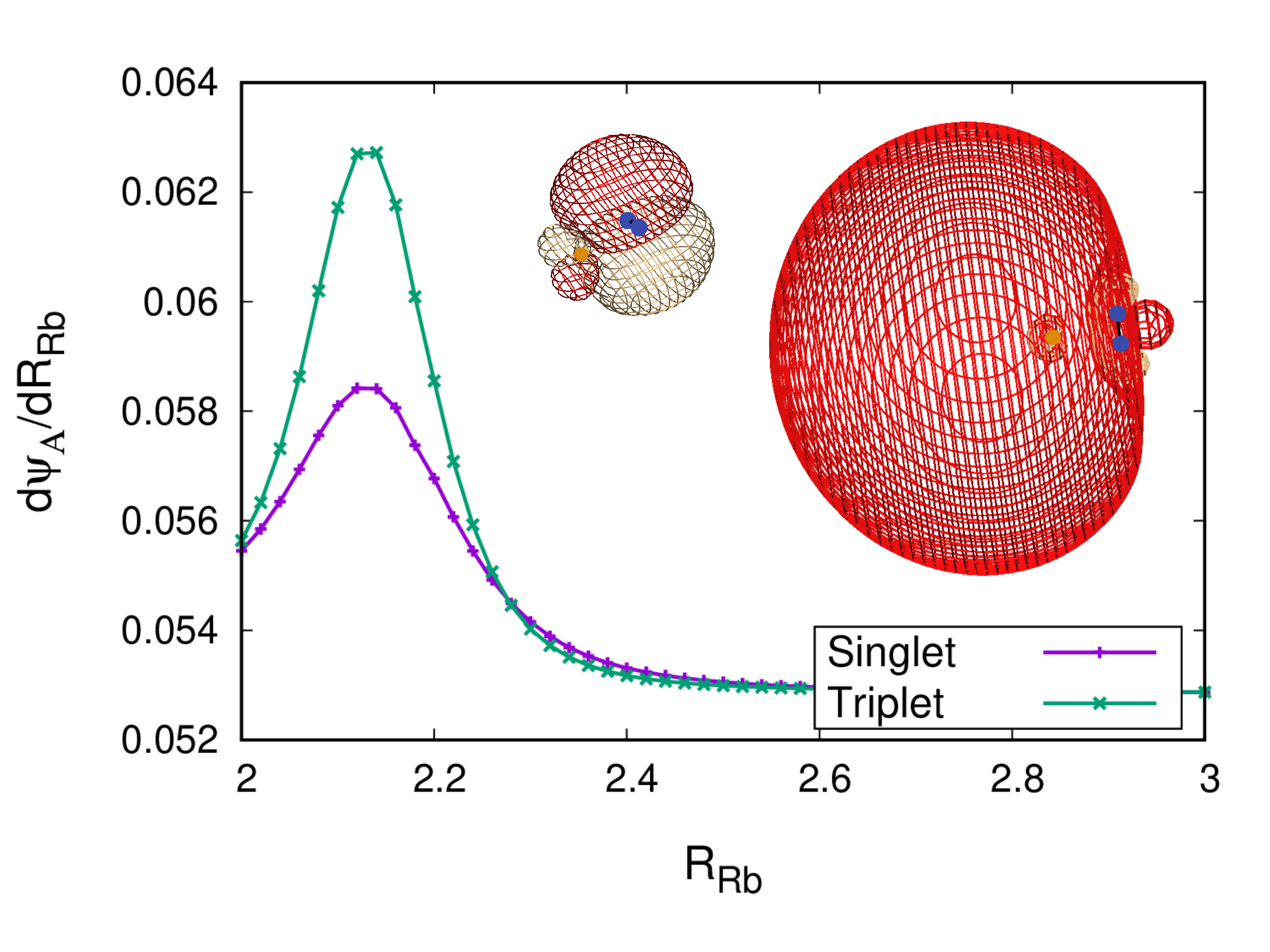}
\end{minipage}
\begin{minipage}[]{10pc}\caption{\label{fig_dpsi} Derivative of the wave function along the $R$ coordinate for Rb-C$_2^-$ and shape of the HOMO for the singlet (left) and triplet (right) state. Note that the triplet state exhibits a strong dipole-bound character.}
\end{minipage}
\end{figure}

From the first excited entrance channel, M($^{2}S$)+C$_{2}^{-}$($^{2}\Pi_u$), a crossing with the neutral PES occurs close to the minimum of the well (shown by the gray circle in Figure \ref{fig_PES}). The first mechanism will therefore dominate the detachment process and, given the large rate of electron detachment when the crossing is reached, the rate of associative detachment is likely to be close to the capture rate. The entrance channel  M($^{2}P$)+C$_{2}^{-}$($X\, ^{2}\Sigma^{+}_{g}$) lies even higher in energy and every collision will cause the collisional complex to enter the autodetachment region as well. Since M($^{2}P$) posses a quadrupole moment, the additional quadrupole-charge interaction will increase the capture rate compared to the Langevin one and we find that $k\approx 1\times 10^{-8}$ cm$^3$s$^{-1}$ below room temperature.  
On the other hand, for the lowest entrance channel, M($^{2}S$)+C$_{2}^{-}$($X\, ^{2}\Sigma^{+}_{g}$), no crossing occurs between the singlet or triplet PES of the anion and the neutral PES. As a result, associative detachment can only be triggered by the second mechanism, $i.e.$ non-adiabatic couplings between nuclear and electronic motions. The detachment process from the ground state entrance channel should then be hindered. Differences between the singlet and triplet states should arise as the HOMO of the triplet state is more diffuse and varies more rapidly with $R$, as shown in Fig. \ref{fig_dpsi}.

\subsection{Inelastic Collisions}\label{section_inelastic}

Given that the AED reaction is predicted to be inefficient for collisions of C$_2^-$ in the ground electronic state with Li and Rb atoms, we can investigate the rotational cooling dynamics of C$_2^-$. The close-coupling approach \cite{Arthurs1960} was employed to calculate cross sections and rate coefficients for rotationally-inelastic collisions involving rotational states up to $j=20$. The collision can occur on the singlet or the triplet PES, and the total cross section is given by the weighted sum with a ratio 1:3. 
For both Li and Rb, the cross sections are similar for the two spin symmetries, while the rate coefficients are larger for collisions involving Li than for Rb atoms \cite{Kas2019a}.

\section{Cooling model}\label{section_cooling}

Based on the preceding results and discussion, we can develop a model to describe the simultaneous translational and rotational cooling of C$_2^-$ through collisions with co-trapped, laser-cooled Li and Rb atoms.

The rotational cooling will be determined by the evolution of the population of the rotational states, which depend on the state-to-state inelastic rate coefficients:
\begin{equation}\label{eq_pop_evolution}
\frac{dp_j(t)}{dt}=-n_a \sum_l  k_{jl}(T)p_j + n_a \sum_l k_{lj}(T)p_l 
\end{equation}
where $n_a$ is the density of laser-cooled atoms, $p_j$ is the population of the rotational level $j$ of C$_2^-$, and $k_{jl}$ is the rate coefficient for inelastic collisions from level $j$ to level $l$. The rate coefficients depend on the translational temperature $T$. 
The translational cooling of the anions is determined by elastic collisions and can be modeled by considering hard-sphere interactions as a function of the number $N$ of collisions as $dT/dN=-(T-T_a)/\kappa$, where $T_a$ is the temperature of the atoms and $\kappa=(m_i+m_a)^2/(2m_im_a)$ \cite{deCarvalho1999}. This equation can be recast as
\begin{equation}
\frac{dT}{dt}=-\frac{R(T)}{\kappa}(T-T_a)
\end{equation}
in terms of the collisional rate $R(T)=n_a\sigma_{\mathrm{el}}v$, where the elastic cross section $\sigma_{\mathrm{el}}$ and the velocity $v$ depend on the temperature, while the density of the atoms is assumed to be constant. This equation is solved assuming an initial temperature $T_0=300$~K.

A general characteristics of rf traps is that ions undergo micromotion, leading to non-thermal energy distributions \cite{Chen2014} and  limiting the minimal temperature of sympathetically cooled ions \cite{Cetina2012}. Several approaches can be employed in order to reduce the amount of micromotion and the residual heating, among which using systems with a small atom-ion mass ratio $\xi=m_a/m_i$ \cite{Haze2018} or increasing the pole order of the trap. For instance, in the case of OH$^-$ ions in an octupole trap cooled by ultracold Rb atoms, temperatures of less than 0.1 K should be reachable \cite{Holtkemeier2016} despite the large value of $\xi$.
Without taking into account the complete trap geometry, the effect of the non-thermal velocities can be investigated by computing the inelastic rate coefficients using a Tsallis distribution instead of a Maxwellian one. The characteristics of this distribution is a power-law tail that depends on a parameter $q>1$ as well as on the pole order of the trap \cite{Holtkemeier2016,Meir2016}. The Maxwell-Boltzmann distribution is recovered for $q\rightarrow 1$ and large pole orders.

Another factor limiting the feasibility of cooling is the presence of excited states of the ultracold atom due to the MOT laser. This leads to two additional loss channels: inelastic collisions M($^2P$) + C$_2^-$ $\rightarrow$ M($^2S$) + C$_2^-$ in which the energy released is larger than the trap depth, and associative detachment M($^2P$) + C$_2^-$ $\rightarrow$ MC$_2$ + $e^-$.
As discussed in Section \ref{section_collisions}, the rate coefficient for associative detachment from the excited state M($^2P$) is expected to be close to the value provided by capture theory. The impact of the excited states of the ultracold atoms can be somewhat mitigated by using a dark spontaneous force optical trap (dark SPOT), allowing the population of the excited state to remain below 10\%.

\begin{figure}[]
\centering
\hspace{-4pc}%
\begin{minipage}{.55\textwidth}
\includegraphics[width=1\textwidth]{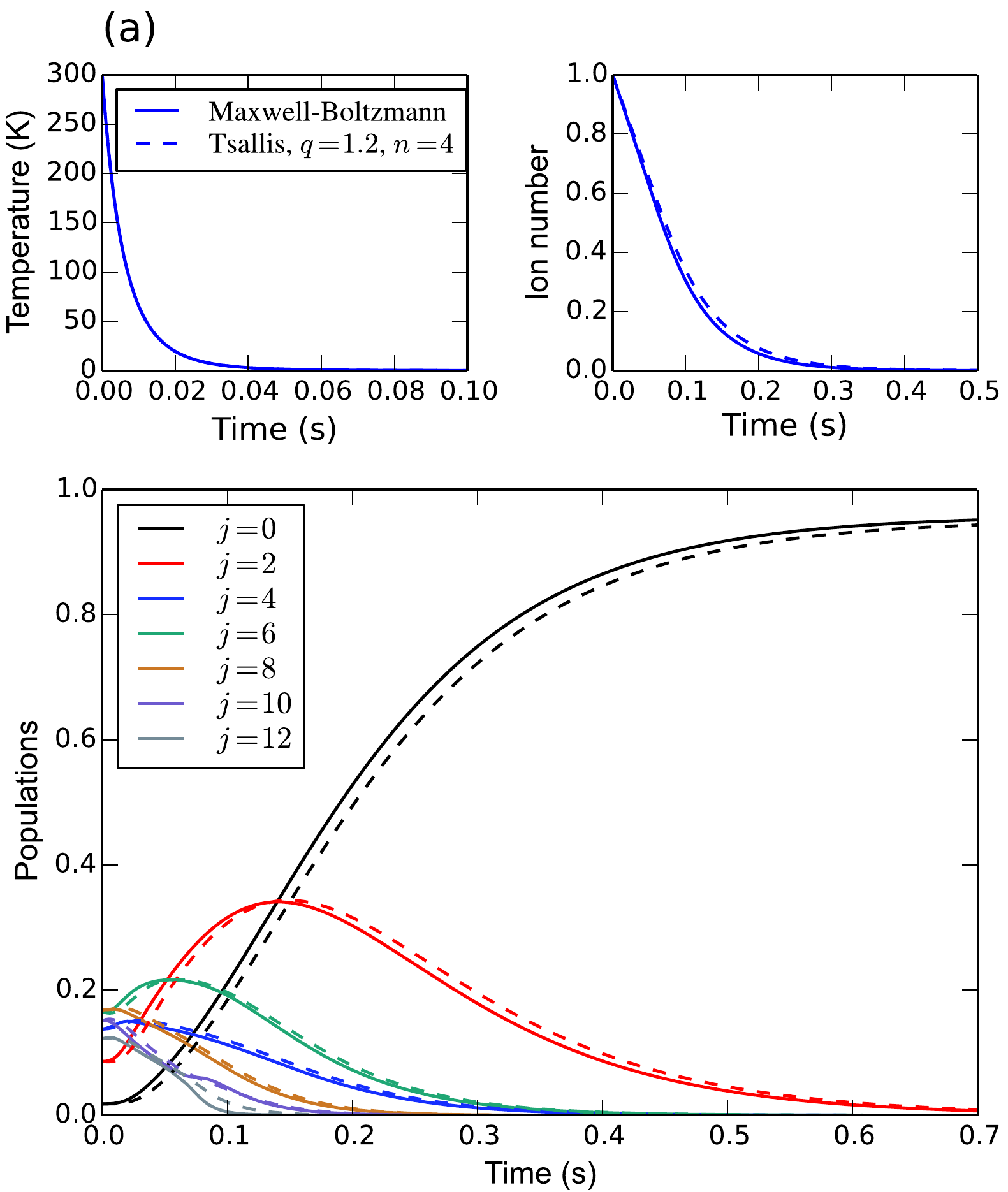}
\end{minipage}%
\begin{minipage}{.55\textwidth}
\includegraphics[width=1\textwidth]{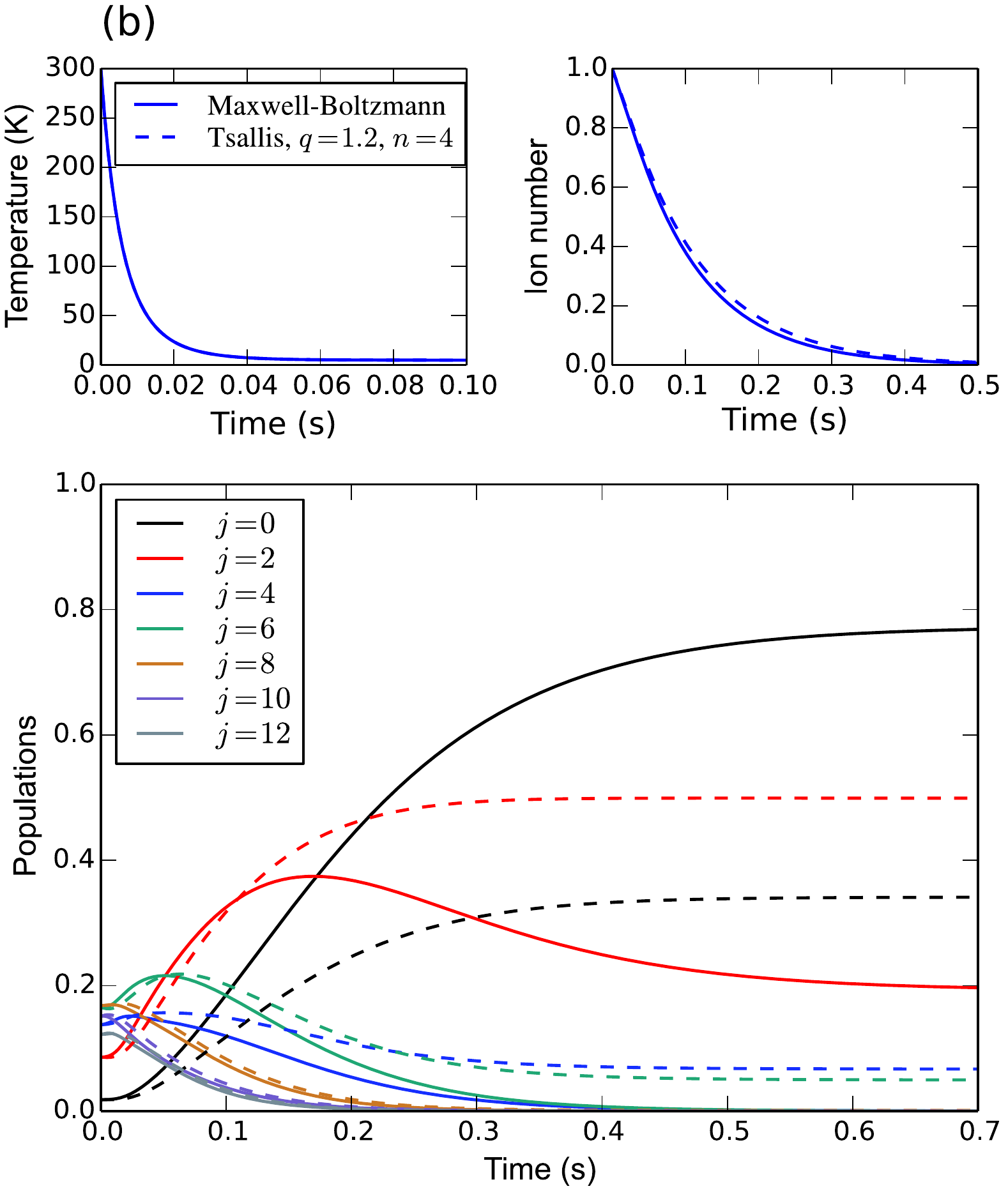}
\end{minipage} 
\caption{\label{fig_cooling}
Evolution of the translational temperature, of the number of ions, and of the rotational populations as a function of time. (a) $T_a=0.1$~K; (b) $T_a=5$~K. Full lines: results obtained with a Maxwell-Boltzmann distribution of velocities; dashed lines: results obtained with a Tsallis distribution with parameter $q=1.2$ for an octupole trap. The atom density is $n_a=10^{10}$cm$^{-3}$ and the fraction of excited atoms is 5\%.
}
\end{figure}

The results of our simulations are shown for Rb-C$_2^-$ in Fig. \ref{fig_cooling}. For two atom temperatures ($T_a=0.1$~K and $T_a=5$~K) and two distributions of velocities (Maxwell-Boltzmann or a Tsallis distribution with parameter $q=1.2$ in an octupole trap), we compare the evolution of the translational temperature, of the number of ions, and of the rotational populations. The atom density is fixed to $n_a=10^{10}$cm$^{-3}$ and the fraction of excited atoms is 5\%. We observe that the translational temperature reaches its limit much more rapidly than the rotational temperature. For $T_a=0.1$~K the difference between the two distributions is small, as could be expected, while for $T_a=5$~K the effect is more pronounced. With the Maxwell-Boltzmann distribution the rotational populations approach the value corresponding to a rotational temperature identical to that of the buffer gas since the rate coefficients satisfy detailed balance. For the Tsallis distribution, on the other hand, the power tail leads to a higher population of excited rotational states, and thus to a rotational heating. This heating is more pronounced for low values of the multipole order, as expected (not shown). 

It is worth noting that even with a fraction of Rb($^2P$) of 5\%, the large rate coefficient for the AED reaction still suggests a rapid depletion of the molecular anions from the trap. This depletion occurs on a timescale longer than that of the translational cooling, but shorter than that of the rotational cooling. This will fundamentally limit the possibility of producing internally cold C$_2^-$ anions in collisions with Rb atoms. Increasing the density of the Rb atoms will lead to a faster cooling, but also to more losses due to the associative detachment collisions with the excited atoms.

Finally, for Li atoms the mass ratio $\xi$ is much more favorable so that the heating due to micromotion is reduced. On the other hand, the rate constant for associative detachment from the excited $^2P$ state is similar to that of Rb, so the loss of anions should be comparable.

\section{Conclusions}

The prospects for the sympathetic cooling of molecular anions by collisions with ultracold atoms in hybrid traps have been explored for C$_2^-$. In general, the efficiency of the cooling process is limited by factors such as heating through coupling with the micromotion in the rf trap and reactive collisions. The first concern can be addressed by using systems with low atom/anion mass ratios, high trap orders, or a localized buffer gas \cite{Holtkemeier2016}. 
In order to limit the impact of associative detachment reactions, one general rule would be to choose a closed shell AB$^{-}$ species with a large electron affinity, and systems other than OH$^-$ and C$_2^-$ have been proposed \cite{Tomza2017,Kas2019b}.  However, the position of the associative detachment region depends on the atomic collision partner and must be investigated using accurate \textit{ab initio} methods. In addition, the presence of excited states of the atoms in the MOT leads to a loss as these atoms will react with the anions on a timescale similar to that of the internal cooling. A complementary approach to cool anions is to use helium buffer gas cooling \cite{deCarvalho1999,Gianturco2018a}, although in that case the temperatures are limited to a few kelvins. The dynamics of buffer-gas cooling can also be investigated based on the calculated inelastic cross sections \cite{Schullian2015,Doppelbauer2017}.

\ack
We acknowledge financial support from the Fonds de la Recherche Scientifique de Belgique-FNRS through the IISN 4.4504.10 grant. Computational resources have been provided by the Shared ICT Services Centre of the Universit\'e libre de Bruxelles.

\section*{References}

\providecommand{\newblock}{}

\end{document}